\documentclass[,final ] {aipproc}

\layoutstyle{8x11single}

\usepackage{latexsym}
\usepackage{epsfig}

  \usepackage{graphicx}
\usepackage{caption}
\usepackage{subcaption}

\usepackage{overpic}
\usepackage{subfig}
\usepackage{tabularx}

\makeatletter
\newcommand{\thickhline}{%
    \noalign {\ifnum 0=`}\fi \hrule height 1pt
    \futurelet \reserved@a \@xhline
}
\newcolumntype{"}{@{\hskip\tabcolsep\vrule width 1pt\hskip\tabcolsep}}
\makeatother

\def\pf{\overline{p_{\rm F}}}
\def\pb{\overline{p_{\rm B}}}

\def\pt{p_{\rm T}}
\def\av#1{\langle #1 \rangle}
\def\bcor{b_{\rm corr}}

\def\Rstr{R_{\rm string}}

\def\betastr{\beta_{\rm string}}
\def\betastrv{\vec{\beta}_{\rm string}}

\begin{document}

\title[Mean transverse momenta correlations in MC toy model]
{Mean transverse momenta correlations
in hadron-hadron 
collisions in MC toy model with repulsing strings}


\author{Igor Altsybeev }{
  address={St. Petersburg State University}
}

\begin{abstract}
In the present work, Monte-Carlo toy model with repulsing quark-gluon strings
in hadron-hadron collisions is described.
String repulsion creates transverse boosts 
for the string decay products, giving modifications
of   observables.
As an example, long-range correlations between mean transverse momenta of particles in two observation windows  are studied in MC toy simulation of the heavy-ion collisions.

\end{abstract}

\maketitle

\section{Introduction}
\label{sec:intro}

Interactions between quark-gluon strings in hadron-hadron collisions are the topic of interest 
for many years. The models of such interactions, however, are mostly phenomenological. For instance, the string fusion model was proposed in \cite{stringFusion1, stringFusion2, stringFusion3}.  It was shown that 
the string fusion phenomenon should lead to modifications of event multiplicity,   transverse momentum spectrum, and to other consequences. 
The string fusion scenario
 was implemented in 
a number of MC models of hadron-hadron collisions
\cite{psm,vk_modelDescr}.

In \cite{abramovsky},   
an attraction and a repulsion of  
chromoelectric tubes in hadron-hadron collisions is discussed. 
It is shown, that   
such interactions should lead to azimuthal asymmetry in the distribution of secondary particles. The following picture is considered:
\begin{enumerate}
\item quark-gluon tubes (strings) have a finite radius.
\item depending on the transverse distance between them, strings may overlap and interact. 
\item strings attract or repel each other in the transverse direction.
\end{enumerate}

In the first part of this proceeding, the toy MC model based on ideas from  \cite{abramovsky} is described.
The motivation for development of such a model comes,
for instance, 
from the results of dihadron correlations
measured in Au-Au collisions in STAR \cite{STAR_2012}, 
where patterns of the collective behavior  are observed and a detailed fit of the correlation structures was developed. 
It is interesting to see what would be the "collectivity" 
in the frame of the toy MC model with repulsion strings.
In the second part of the current proceeding, 
the so-called 
mean transverse momentum correlations are extracted 
from the toy model events. This observable  
may be useful to disentangle between string interaction scenarios, for example, between the string repulsion and 
the string fusion.  
Dihadron analysis of the MC toy model data is
presented in 
the same proceedings \cite{QC_constr_on_perc_model}.

\section{Monte-Carlo Toy Model}
\label{sec:toy}

In this section a Monte-Carlo (MC) toy model with repulsing strings 
is described.
The MC model is applicable to different types of hadron-hadron collisions (pp, AA, pA, etc.). \\ \\
{\bf Stage 1. Simulation of hadron-hadron collisions, strings formation.}\\
In this MC model, initial positions of the nucleons in nuclei are generated in accordance to Woods-Saxon distribution (for the $\rm Pb^{208}$, the WS radius is 6.62\,fm and parameter $a=0.546$\,fm). Nucleon core effect is not taken into account  to speed-up computations.
Inside each nucleon, some number of partons is distributed in transverse ($xy$) plane with 2D-Gauss law, with $\sigma_{xy}=0.4$ fm. The {\it mean number of partons} $n_{\rm partons}$ inside nucleons is dependent on a collision energy and is a model parameter.
Interaction between colliding hadrons is implemented at the partonic level:
partons can interact and form a $string$,
 if the distance between partons in $xy$ plane is less then some {\it parton interaction distance} $d_{\rm p}$.
There is a $3\%$ probability for a string to be "hard scattered"
(this number is also a model parameter) -- this is used for  jets and jet-like structures simulation.
All other strings are considered to be "soft" and "long" in rapidity, occupying rapidity range  $y\in(-4,4)$.
As the result of this model stage, a configuration of strings appears. 
 Parameters of the MC model are given in Table \ref{tab:toyModelParams}.

\begin{figure}[h]
%\begin{center}
\centering
\begin{overpic}[width=0.55\textwidth]{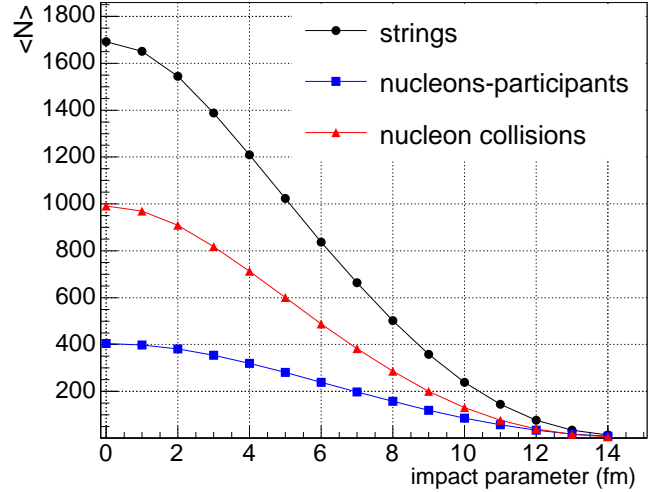} 
          %\put(12,32){\Huge (a)}
\end{overpic}
%\end{center}
\caption{Dependence of the mean numbers of nucleons-participants, nucleon-nucleon collisions and  strings on impact parameter in the toy model.
}
\label{fig:NstrNpartNcoll}
\end{figure}

If at least one parton of the nucleon has interacted, this nucleon is considered as  nucleon-participant.
Fig.\,\ref{fig:NstrNpartNcoll} shows mean number of nucleons-participants, nucleon-nucleon collisions and average number of produced strings as a function of impact parameter. One can recognize  similar dependences 
in (nucleon-level) Glauber MC calculations.

\begin{table}[h]
\centering
\begin{tabular}{p{5cm} p{3cm} | p{2cm} p{2cm} p{2cm}}
\thickhline
{\bf Parameter } & {\bf Notation} &\multicolumn{3}{c} {\bf Values} \\ \thickhline
average number of partons & $\av{n_{p}}$ & 15   \\ \hline
parton interaction $xy$ distance & $d_{\rm p}=2r_{\rm p}$ & 0.4 fm & \\ \hline
string energy density & $\lambda$ & 1.0 GeV\\ \hline
string radius &$\Rstr$ & 0.25 fm & 1 fm  & 2 fm\\ \hline
string overlap energy density & $\lambda_1$ & 0.2 GeV & 0.002 & 0.0001 GeV\\ \hline
\end{tabular}
\caption{ 
Parameters of the %strings in 
MC toy model.
}
\label{tab:toyModelParams}
\end{table}

\noindent {\bf Stage 2. Repulsion of the strings.}\\
At the next step of the system evolution,  "soft" strings  interact with each other.
In the current MC model, interaction 
manifests itself as a repulsion.
The following 
repulsion mechanism is considered \cite{abramovsky}.
Each string has some {\it effective interaction radius} $\Rstr$, with corresponding string {\it effective transverse area} $S_{\rm string}=\pi \Rstr^2$.
A string
which is far apart from the others 
has energy density per unit length $\lambda$.
Two completely overlapped strings have energy density $2\lambda + 2\lambda_1$,
while  density of partial  overlapping is $2\lambda + 2\lambda_1\cdot S/S_{\rm string}$, 
where $S$ is the area of the overlap (i.e. it is assumed that effectively strings are "black discs" in the transverse plane).
Here $\lambda_1$ is the  energy density excess due to overlapping.
When two strings interact in such a way, each string acquires transverse momentum 
\begin{equation}
p_{\rm T\,string}=l\sqrt{(\lambda + \lambda_1  S/S_{\rm string})^2-\lambda^2}\ .
\end{equation}
After vector sum of all instantaneous momentum kicks, a string gains corresponding transverse relativistic boost   
$\betastrv$. 
For different values of the model parameter $\Rstr$, $ \lambda_1$  is manually adjusted in such a way that 
event-averaged string boost
$\av{\betastr}\approx 0.65$ keeps constant
in the most central collisions (this value is arbitrary chosen).

Doubled string radius $2\Rstr$ can be interpreted as the effective interaction distance between strings  in the transverse plane. 
For example, it may be understood as a "mean  path" of the string  before it hadronizes.

\begin{figure}[h]
%\begin{center}
\centering
$\begin{array}{ccc}
\includegraphics[width=0.45\textwidth]{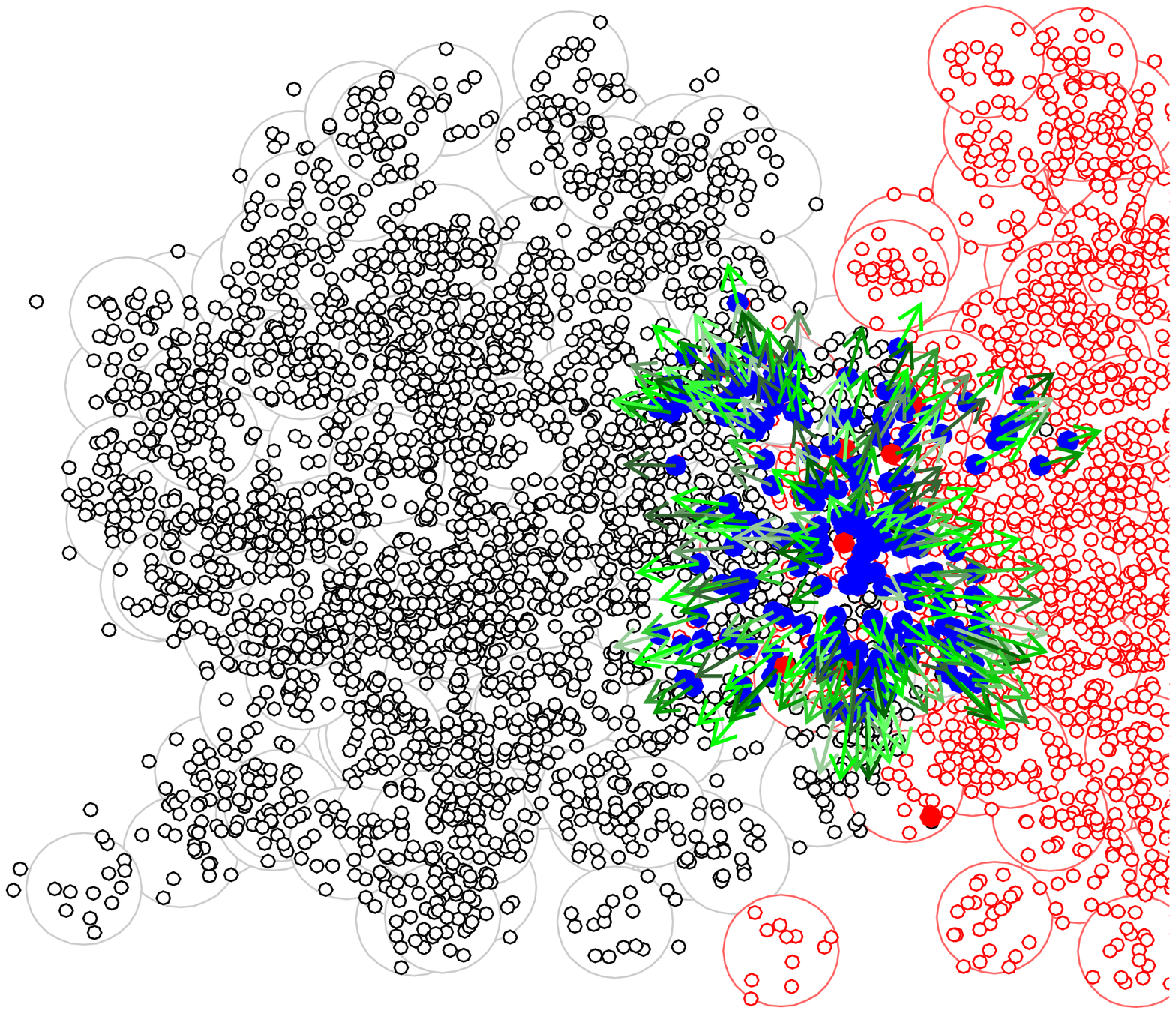} 
&
\includegraphics[width=0.45\textwidth]{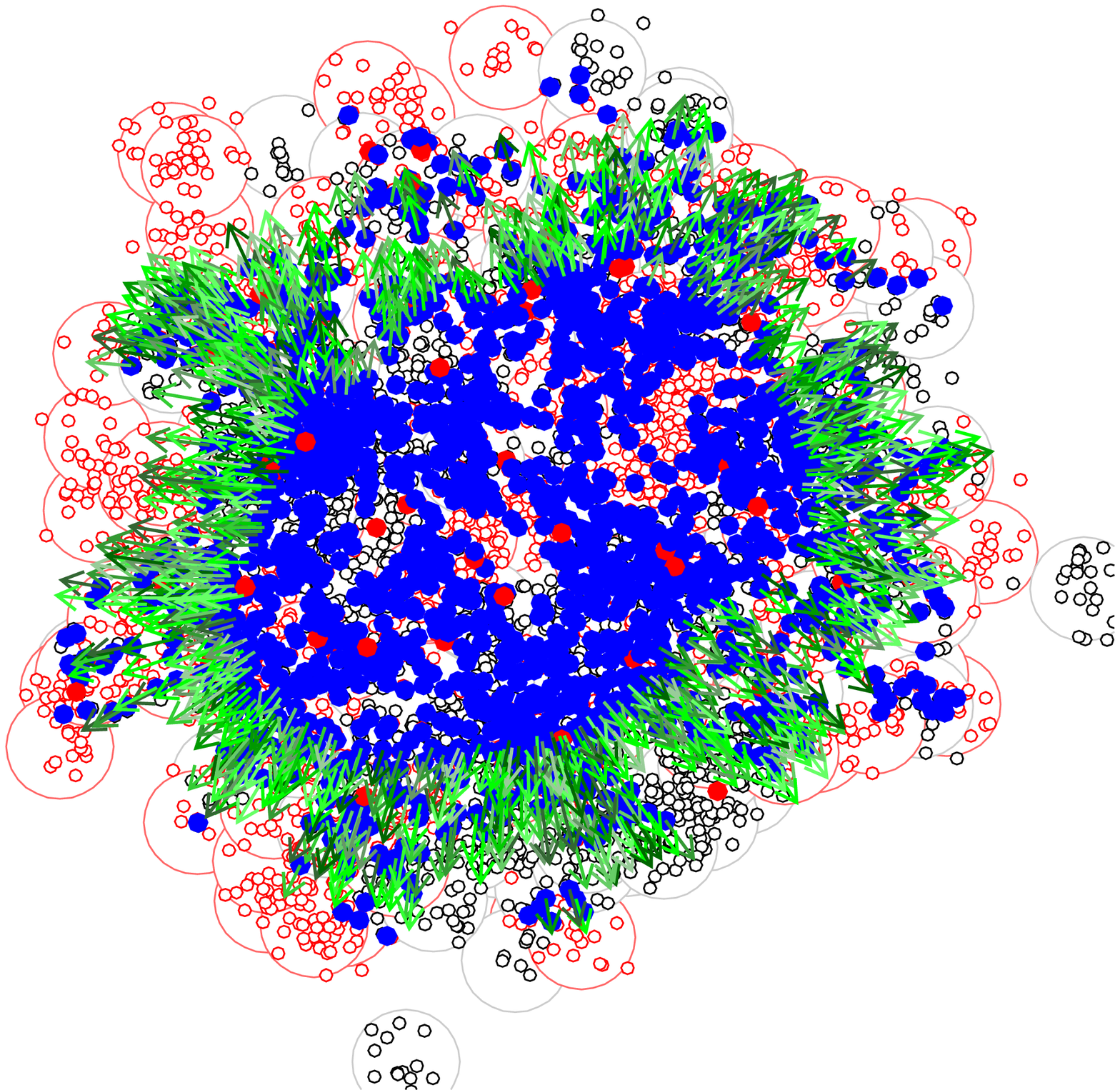} 
\end{array}$
%\end{center}
\caption{
Examples of peripheral (left) and central (right) events in MC toy model, 
with impact parameter $b=12$ and $0$ fm, respectively.
}
\label{fig:eventView} 
\end{figure}

At this stage of the event evolution, before hadronization of the strings, some figures can be plotted in order to understand the model behavior.
Fig.~\ref{fig:eventView} shows cartoons for peripheral (left) and central (right) events.
Colliding nuclei and the structures inside them are  in red for one nucleus and  in black colors for the second. Nucleons are shown as large  circles,  smaller circles inside illustrate partons (the radii of the circles reflect the actual model parameterization). Blue full circles denote positions of the strings, and each green arrow shows the direction and magnitude of the gained Lorentz boost for one string (some threshold on minimum arrow length is applied in order to hide too short arrows). Red full circles are for "hard scattered" strings, which do not participate in the repulsion calculation.

\begin{figure*} 
\centering
$\begin{array}{ccc}
\begin{overpic}[width=0.49\textwidth]{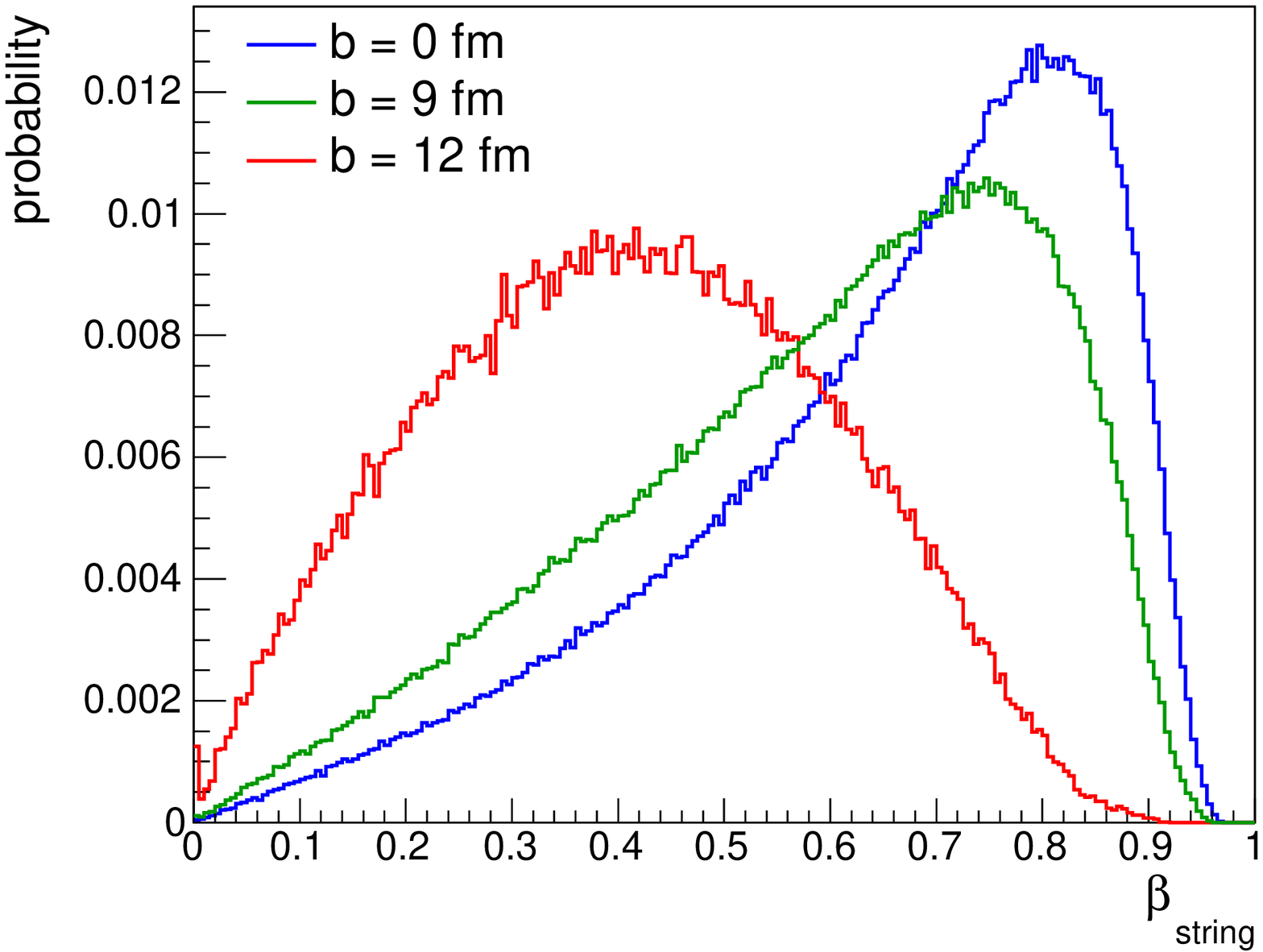} 
          \put(44,68){string radius = 1 fm}
\end{overpic}
&
\begin{overpic}[width=0.49\textwidth]{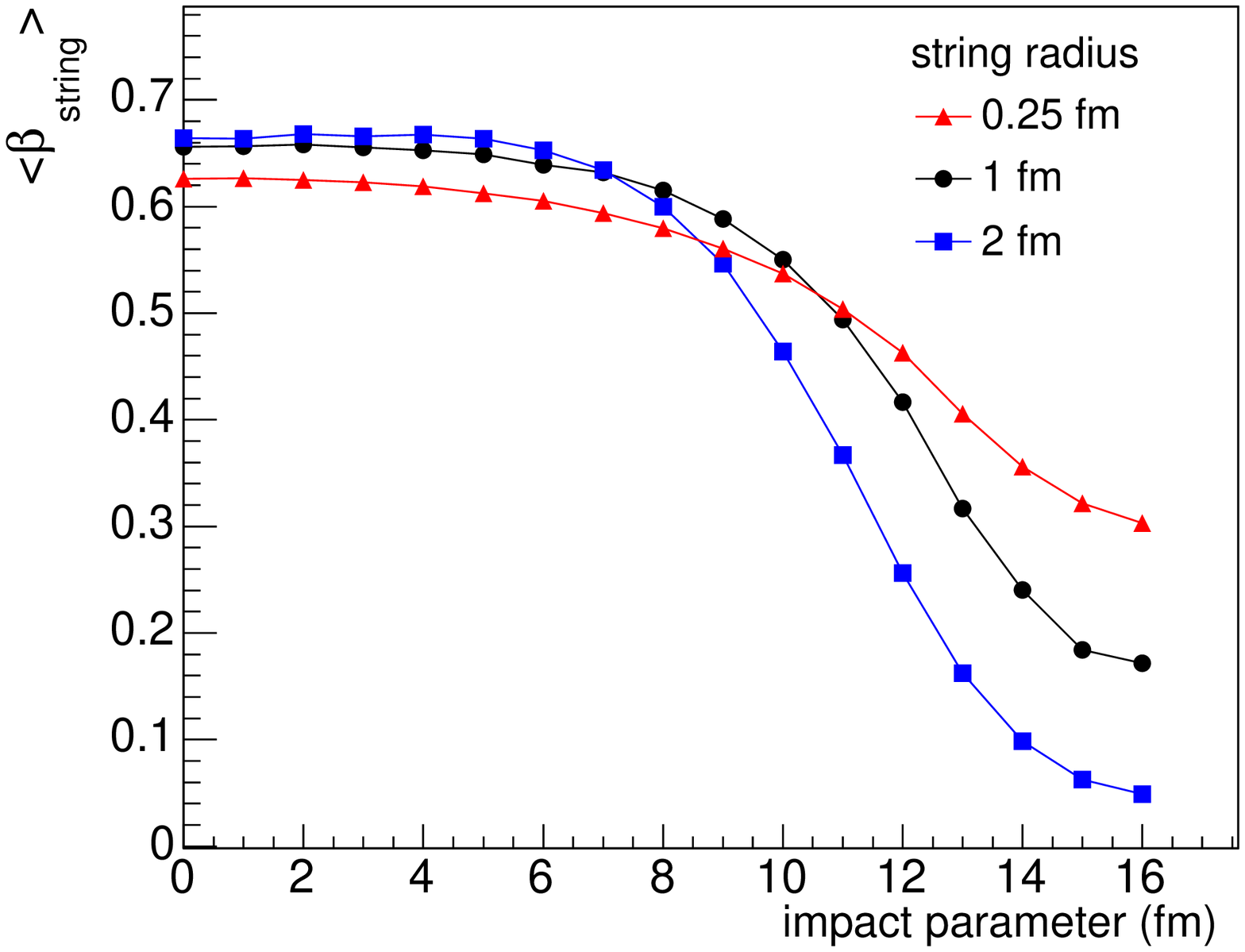} 
          %\put(12,32){\Huge (a)}
\end{overpic}
\end{array}$
\caption{
String $\beta$-boost distribution for different impact parameters.
}
\label{fig:betaStrings} 
\end{figure*}

The probability distribution of string boosts $\betastr$ 
is shown in Fig.\ref{fig:betaStrings} (left)  for several values of impact parameter~$b$. The shape of distribution  changes with $b$ -- in more central events, strings in general gain larger boosts.
Fig.~\ref{fig:betaStrings} (right) shows average value $\av{\betastr}$ as function of $b$ for simulations
with different string interaction radii $\Rstr$.
For all three   radii, some transition region from peripheral to central $b$ can be seen (more "step-like" behavior in case of  $\Rstr$=2 fm), and
in more central events $\av{\betastr}$ is nearly constant.

{\noindent\bf Stage 3. Hadronization of the strings.}\\
Naive hadronization mechanism implemented in this toy model is the following.
In its rest frame, each string breaks  at several points, 
and in each break-point a~pair of quarks appears.
The number of break-points is proportional to the string length, with some small random fluctuations.
A random transverse momentum
 distributed  exponentially ($\sim e^{-\alpha\pt}$) and
random azimuthal direction are assigned to one quark in a pair; the second quark flies out with the same modulus of $\pt$, but in opposite $\varphi$-direction. Each piece of the broken string has two quarks at its ends and  is treated as a "particle" (meson). Namely, 
in this toy model all "particles" originated from the string 
are assigned   
to be exclusively $\rho$-mesons. 
This assumption simplifies the model, reflecting at the same time the fact that number of $\rho$ mesons in high-energy AA collisions is quite significant.
The transverse momentum of $\rho$ is calculated as a vector sum of transverse momenta of its two quarks. 
The probabilities for the charge of each $\rho$-meson  to be positive, negative or neutral are 25\%, 25\% and 50\%, respectively, 
and the $\rho$ mass is assumed to be 775~MeV with Gaussian smearing with $\sigma=80$~MeV.
The only decay channels are  
$\rho^+\rightarrow \pi^+\pi^0$,
$\rho^-\rightarrow \pi^-\pi^0$ and
$\rho^0\rightarrow \pi^+\pi^-$, 
with the relativistic decay kinematics implemented.

Transverse Lorentz boost $\betastr$ of each $\rho$ meson is performed at rapidity of $y=0$. After that, random $y$, uniformly distributed along the string length, is assigned to this $\rho$. This approach is quite a rough one, since transverse Lorentz boosts for particles with large $|y|$ 
should significantly  modify $dN/dy$ distribution.
However, the distribution of the string boost along $y$ 
is unknown, thus any reasonable model assumptions are allowed.
%Decay products ...
More sophisticated algorithm of string fragmentation 
can be found, for example, in the VENUS model \cite{Werner}.

{\noindent\bf A final state of the event.}\\
The key feature of the toy model is that
{\it in the laboratory frame, all the particles originated from one string  are boosted by the factor $\betastrv$.}
So,
in the final state of a toy event, there is a number of 
$\pi$-mesons, which can be "detected".

\begin{figure*} 
\centering
$\begin{array}{ccc}
\begin{overpic}[width=0.5\textwidth, clip=true, trim=0 0 0 0]{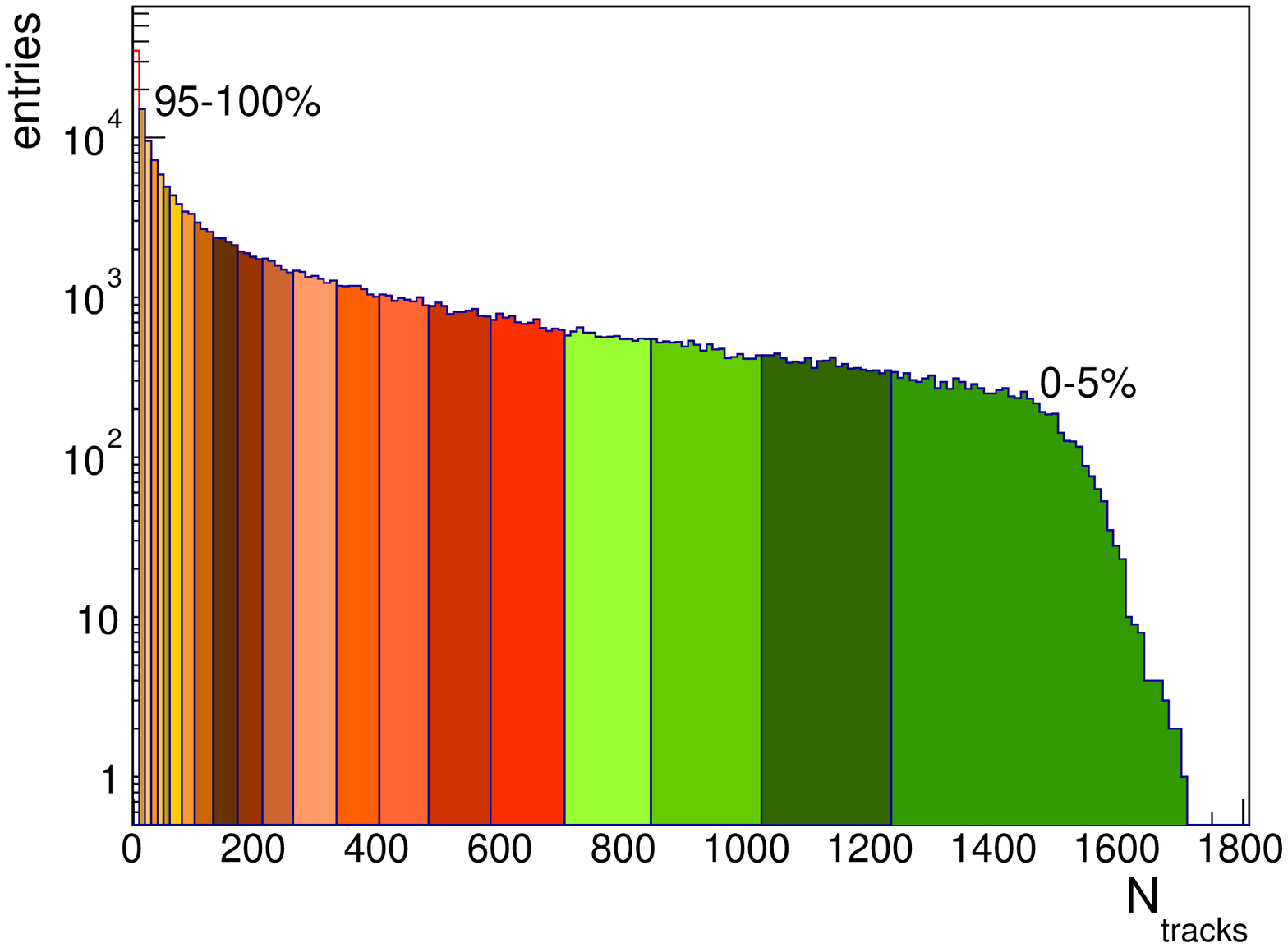} 
          %\put(12,32){\Huge (a)}
\end{overpic}
&
\begin{overpic}[width=0.5\textwidth]{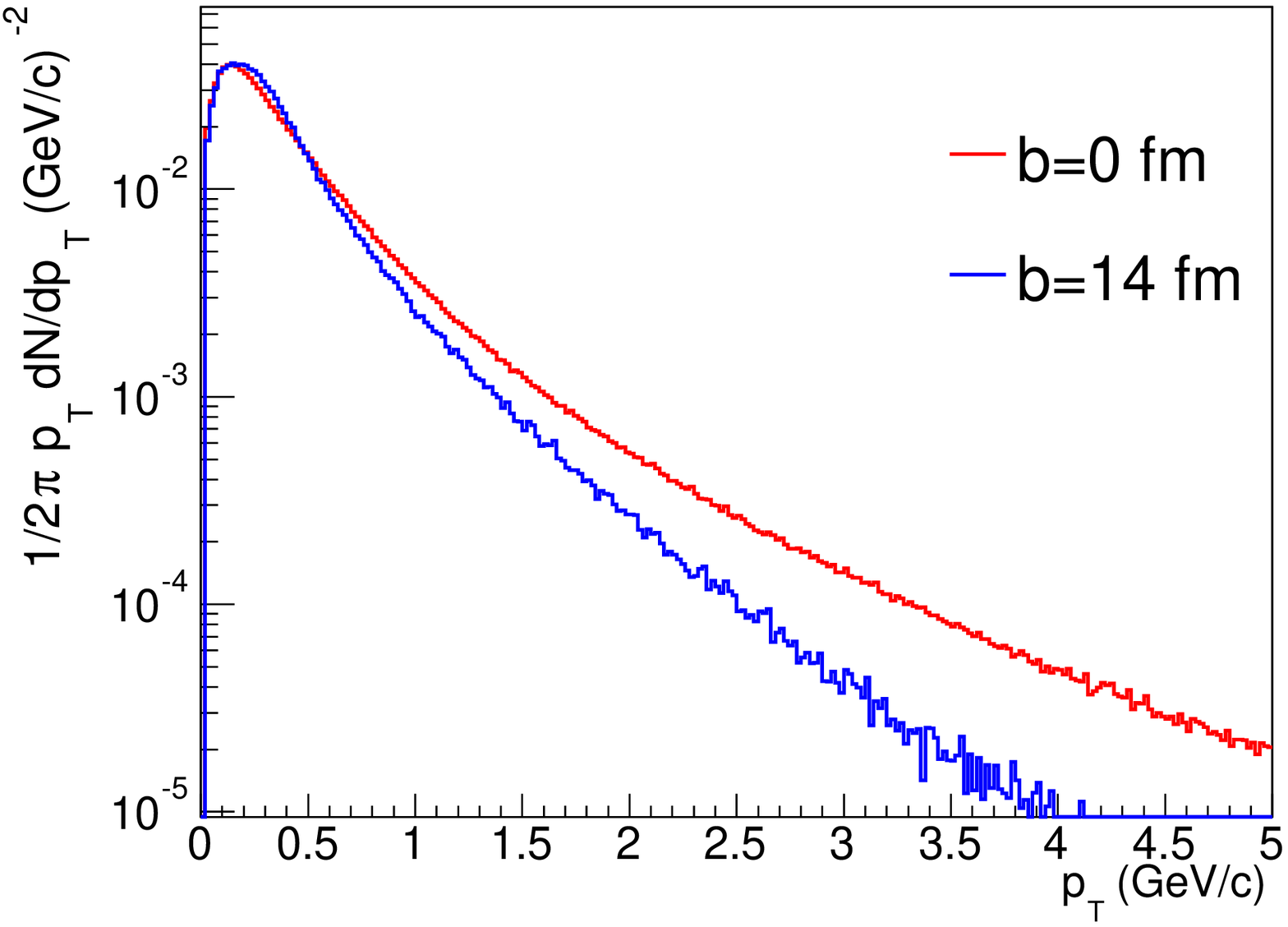} 
          %\put(12,32){\Huge (a)}
\end{overpic}
\end{array}$
\caption{
Multiplicity distribution divided into 20 multiplicity classes (5\% each),
obtained in the toy model (left pad).
Pions $\pt$ spectra in toy events for two impact parameters $b=0$ and 14 fm (right pad).
}
\label{fig:multClasses} 
\end{figure*}

Fig.~\ref{fig:multClasses} (left) shows charged particle multiplicity distribution
in pseudorapidity range $|\eta|<0.5$ %(DENSITY PER ETA UNIT!), % $dN_{\rm ch}/d\eta$, 
obtained in the toy model.
In most central events, the multiplicity is $\sim 1600$, 
which is similar to central
Pb-Pb collisions at LHC at $\sqrt{s_{\rm NN}}=2.76$~TeV.
As it is usually done in A-A experiments like STAR and ALICE,
the multiplicity distribution is divided into "multiplicity classes", which are considered to reflect event centralities. These classes are used below in 
the event analysis. Right pad of the Fig.~\ref{fig:multClasses} 
shows $\pt$-spectra of pions in the toy model. The spectra didn't adjusted to fit real data, however, qualitative  behavior of the spectra for central and peripheral events is captured.

\section{Analysis of the toy model events}
\label{sec:corr}

\subsection{Dihadron correlations}

\begin{figure}[h]
%\begin{center}
\centering
$\begin{array}{ccc}
\begin{overpic}[width=0.5\textwidth]{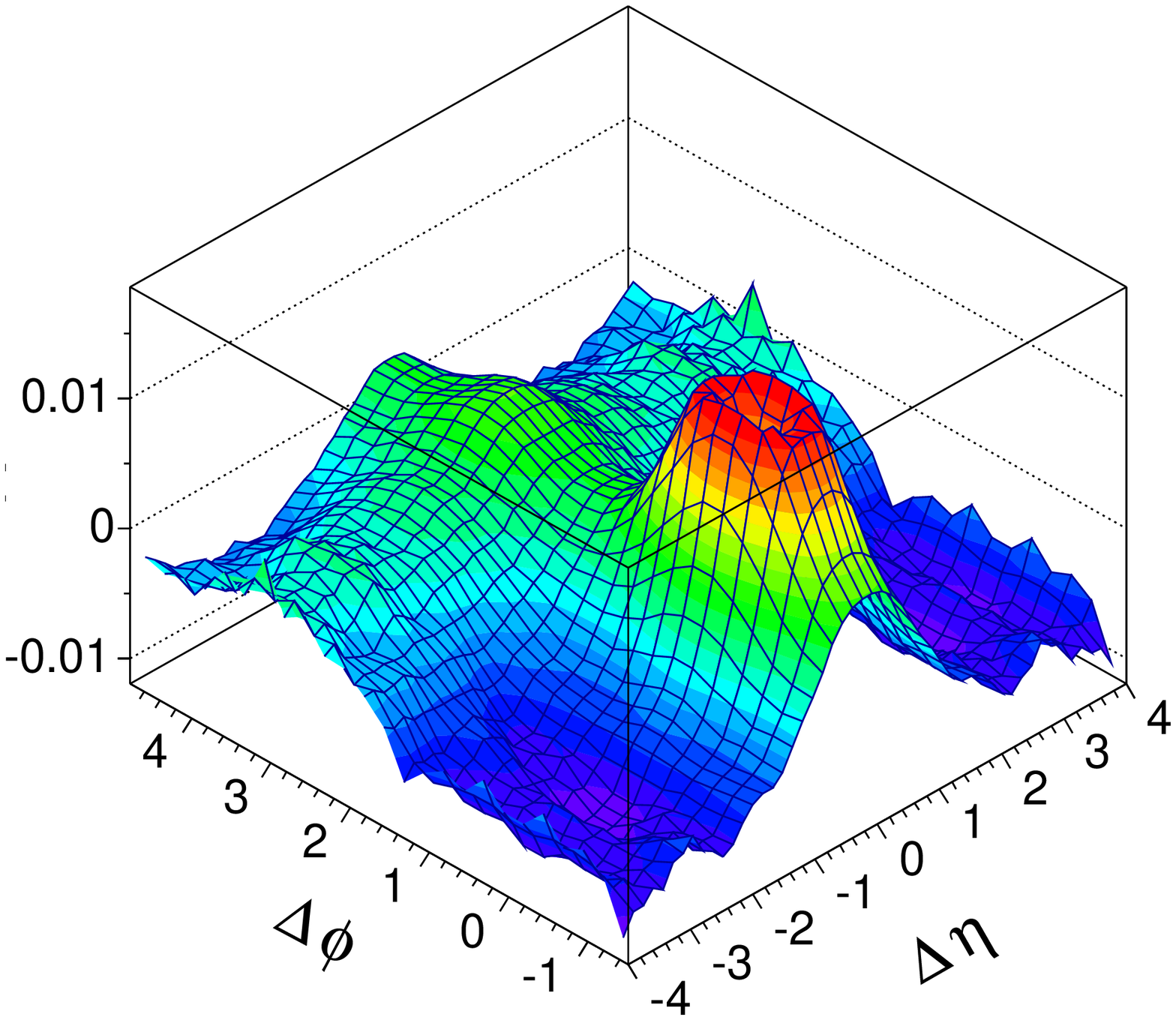} 
	\put(7,74){  peripheral }
\end{overpic}
&
\begin{overpic}[width=0.5\textwidth]{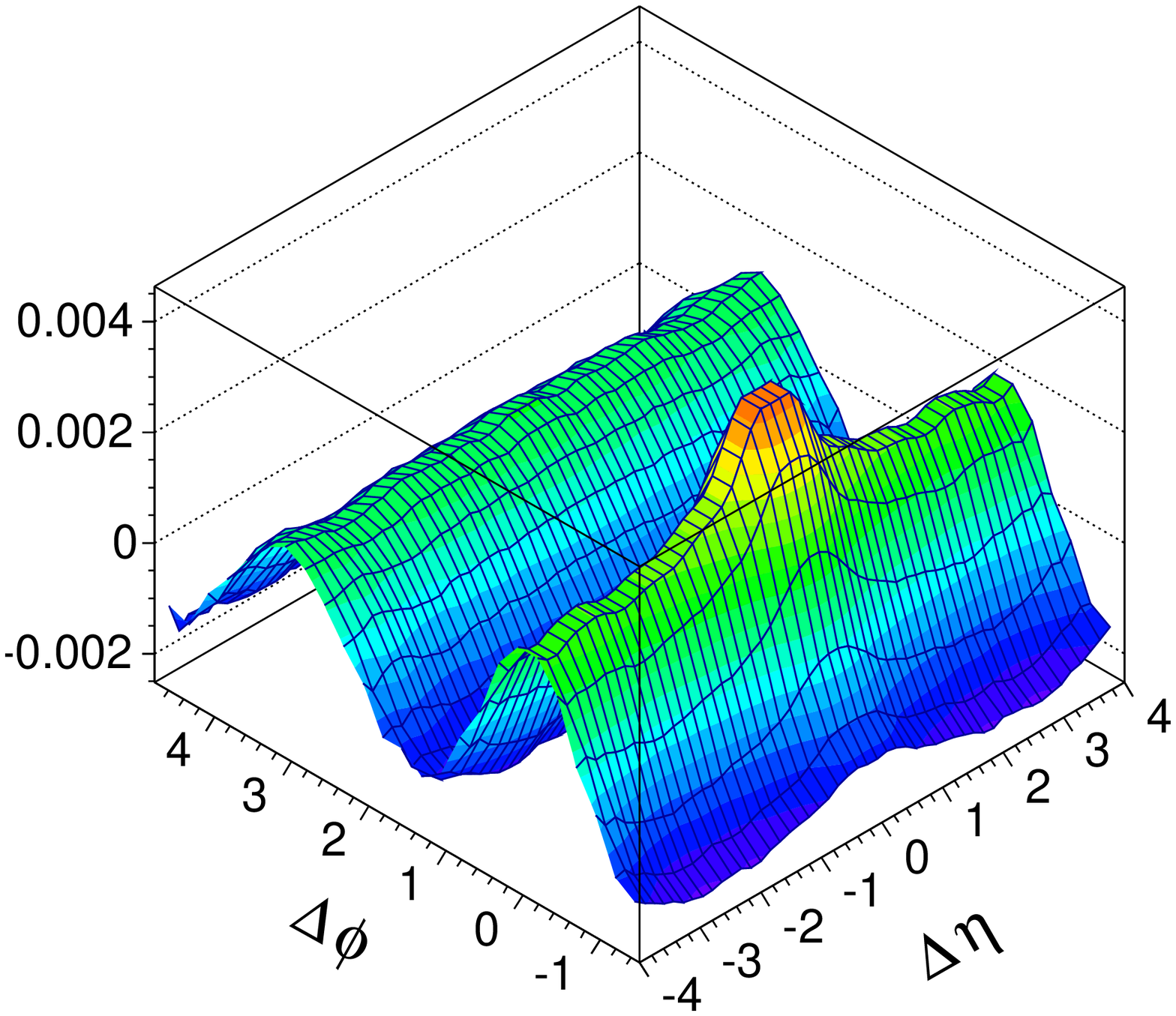} 
	\put(9,74){  central }
\end{overpic}
\end{array}$
%\end{center}
\caption{
Example of dihadron correlation functions obtained in the toy model
for peripheral events (left) and for central events (right).
The dihadron analysis method, used here, is the same as, for example, 
in \cite{cms_dihadron}.
Charged particles with $\pt>0.15$ GeV/{\it c} in $|\eta|<2$ are analysed.
}
\label{fig:dihadron}
\end{figure}

A set of different observables can be chosen %suggested 
to study consequences of the string repulsion, %. Among them are 
such as mean transverse momentum $\pt$,  dihadron correlations, harmonic decomposition analysis, etc. 
As an example,  Fig. \ref{fig:dihadron} shows  results of dihadron analysis 
for peripheral and central Pb-Pb collisions, where tracks  with $\pt>0.15$ GeV/{\it c} are selected.
In peripheral events, a "ridge"-structure along $\Delta\varphi$ is visible, which is due to the $\rho^0$  decays. In central events,
the well-known away- and same-side ridges, which appear along $\Delta\eta$,
can be interpreted as the "elliptic flow" which emerges in the toy system of the repulsing strings. More dihadron results from the MC toy model and the discussion can be found in the contribution called "Constraints on the percolation model from anomalous centrality evolution of two-particle correlations"
in the same proceedings.

\subsection{Mean transverse momentum 
correlations} 

In the current proceeding, the analysis is focused on
 long-range correlations 
obtained with the toy model. The long-range correlations are studied between observables in two different and significantly separated rapidity and azimuthal intervals 
$\eta_{\rm F}$ and $\eta_{\rm B}$, 
which are conventionally refereed as forward (F) and backward (B) rapidity windows.
A correlation coefficient $\bcor$ is defined as the response $\av{B}_{\rm F}$ to the deviation of the quantity F from its expectation value $\av{F}$.
In this work, quantities B and F are chosen to be {\it the average transverse momenta} of all particles produced in a given event in, respectively, the backward and the forward pseudorapidity windows:
\begin{equation}
\label{ptFptB}
F\equiv \overline{\pt}_F= \frac{\sum_{j=1}^{n_F}\pt^{(j)}}{n_F} ,\ 
B\equiv \overline{\pt}_B= \frac{\sum_{i=1}^{n_B}\pt^{(i)}}{n_B} .
\end{equation}

The correlation coefficient 
\begin{equation}
\label{bcor}
b_{\rm corr}^{p_T-p_T}
\equiv\bcor= \frac{\av{FB}-\av{F}\av{B}} {    \av{F^2} - \av{F}^2}
= \frac{\av{\pf\  \pb}-\av{\pf}\av{\pb}} {    \av{\pf^2} - \av{\pf}^2} \ .
\end{equation}

\begin{figure*} 
\centering
\resizebox{0.65\textwidth}{!}
{
\begin{overpic}{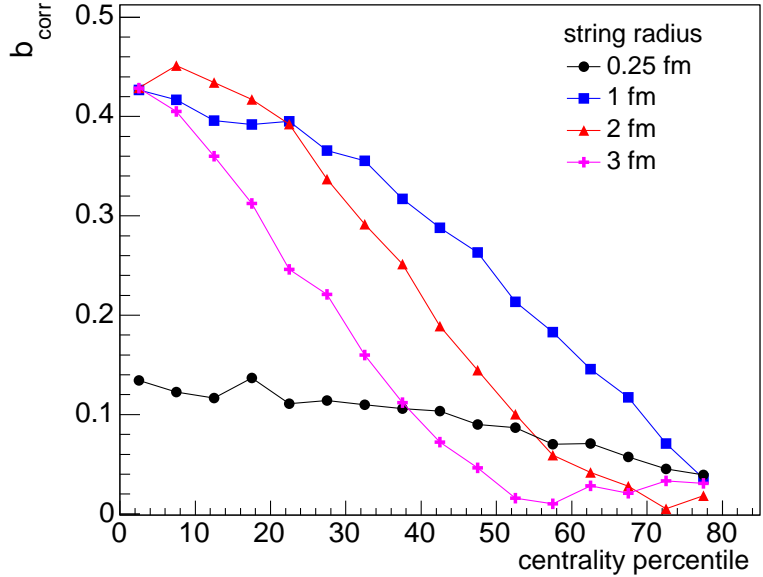} 
          %\put(12,32){\Huge (a)}
\end{overpic}
}
\caption{
Correlation strength  $b_{\rm corr}^{p_T-p_T}$    in centrality classes of the width 5\%, obtained with the MC toy model for Pb-Pb collisions several values of $\Rstr$ (samples of 500k events each). The FB $\eta$-windows are
(-0.8, 0) -- (0, 0.8),  charged particles in $\pt$ range 0.2--2.0 GeV/{\it c} are analysed.
}
\label{fig:ptpt} 
\end{figure*}

Fig. \ref{fig:ptpt} shows   $b_{\rm corr}^{p_T-p_T}$  
in Pb-Pb collisions
in the multiplicity classes of the 5\% width (these classes are defined   
in the left pad in the Fig.~\ref{fig:multClasses}). The model parameters of the  simulation are the same as described before (see 
Table \ref{tab:toyModelParams}).
Forward and backward $\eta$-windows are
(-0.8, 0) and (0, 0.8), respectively.
The values of  $\bcor$ are positive and grow with centrality
to $\approx 0.4$ units for $\Rstr=1,2$ and 3 fm, 
while for small radius
$\Rstr=0.25$ fm  $\bcor$  is significantly lower.
One can notice also, that the growth of $\bcor$
with centrality takes off from zero  earlier for smaller
string radii: for $\Rstr=1$ fm the increase starts already at $\approx 80\%$ of the centrality,
while 
for $\Rstr=3$ fm it happens 
only at centrality $\approx 50\%$.

Long-range correlations  raised due to another possible string interaction mechanism -- string fusion - are studied 
with  MC model \cite{vk_modelDescr} based on color-dipoles for Pb-Pb \cite{vk_PbPb} and for p-Pb collisions \cite{vk_pPb}.

\section{Conclusions}

The MC toy model for hadron-hadron is built,
where quark-gluon strings are formed as a result of
partonic interactions inside colliding hadrons, and the string repulsion mechanism 
is implemented in order to introduce collectivity in the model. 
The toy model is designed to simulate AA, pA and pp collisions.

Behavior of the model with different  parameterizations is explored, it is found that average string boosts in events have a non-linear behavior with centrality and experience saturation for more central events. 
The analysis of the particular observable -- long-range correlations between mean transverse momenta -- is presented. Another kind of analysis 
-- dihadron correlations in the toy model -- is discussed in separate proceeding in the same volume.
The toy model is able to reproduce some experimentally observable features of the azimuthal asymmetry in particle yields. In particular, near-side ridge structure can be obtained.

The model of repulsing strings, however,  have difficulties in description of some observations. For example, the strangeness enhancement can hardly be obtained,
while it is possible  in the string fusion scenario.
One can try to unite several string-based models into one to describe experimental data in a more consistent way.

\begin{theacknowledgments}
Author would like to thank G.~Feofilov, V.~Vechernin, 
V.~Kovalenko and E.~Andronov for useful discussions and advices. 
Author acknowledges Saint-Petersburg State University for the research grant 
11.38.197.2014.
\end{theacknowledgments}

\bibliographystyle{aipproc}   

\bibliography{sample}

\end{document}